\documentclass[11pt]{article}

\usepackage[utf8]{inputenc}
\usepackage[T1]{fontenc}
\usepackage{times}
\usepackage{amsmath,amssymb}
\usepackage{graphicx}
\usepackage{booktabs}
\usepackage{siunitx}
\usepackage{setspace}
\usepackage[hidelinks]{hyperref}
\usepackage{geometry}
\usepackage{longtable}
\title{\textbf{Atmospheric carbon-14 production from neutron leakage in fusion energy systems}}

\author{
B.~J.~Albright$^{1}$\thanks{Email: balbright@lanl.gov}
\and
J.~A.~Mercer-Smith$^{1}$
}

\date{
$^{1}$Los Alamos National Laboratory, Los Alamos, NM 87545, USA\\[6pt]
\today
}

\begin{document}
\maketitle

\begin{abstract}

Neutron-producing fusion systems can generate atmospheric carbon-14 when neutrons leak into nitrogen-containing gas. We use MCNP\textsuperscript{\textregistered}6.2 neutron-transport calculations to estimate the probability that leaked neutrons produce $^{14}$C through $^{14}$N$(n,p)^{14}$C under representative near-ground conditions. For 14.1~MeV deuterium--tritium source neutrons, the conversion probability is 0.25--0.50 across the geometries studied; softer leakage spectra can give larger yields. Scaling this response to a 1~GWe fusion plant shows that percent-level neutron leakage into air would produce an atmospheric $^{14}$C source within a factor of a few of natural global production. At a 2500~GWe fleet scale, limiting fusion-derived radiocarbon to 10\% of the natural source implies a mean atmospheric leakage fraction of order $10^{-6}$. These results provide a screening-level source-term estimate for atmospheric $^{14}$C production from terminal neutron leakage in neutron-producing fusion systems, with particular relevance to architectures containing open ports, beamlines, ducts, or other streaming paths.

\end{abstract}

\section*{Introduction}

Fusion has re-entered the energy discussion with unusual force. Ignition at the National Ignition Facility, together with expanding public and private investment, has strengthened the case that fusion could eventually contribute meaningfully to large-scale electricity supply~\cite{AbuShawareb2022,AbuShawareb2024,IAEA2025,NASEM2021,Callahan2024}. In that discussion, fusion is often described as ``clean'' energy, a description that is broadly correct with respect to prompt radiological hazards and long-lived transuranic waste. It is less complete for neutron-producing fuel cycles because neutrons that escape and enter nitrogen-containing gas can produce atmospheric radiocarbon. This pathway has long been recognized in fusion neutronics~\cite{Scheele1976}, but it has not been treated as a deployment-scale design variable. 
Previous fusion-radiocarbon analyses primarily evaluated material activation, in-vessel inventories, 
coolant or structural source terms, and facility-specific safety questions. 
Here we isolate a different quantity: the atmospheric $^{14}$C yield per neutron that escapes engineered confinement 
into nitrogen-containing gas. This response is then translated into plant- and fleet-scale leakage constraints, 
making the known $^{14}$N$(n,p)^{14}$C pathway a rescalable design and environmental-assessment variable 
for modern fusion deployment.

This study does not estimate the neutron leakage fraction of any specific fusion reactor or facility. Instead, it calculates the atmospheric $^{14}$C response conditional on terminal neutron leakage into nitrogen-containing gas, where ``terminal leakage'' denotes leakage after reactor-specific shielding, port, duct, window, bioshield, or containment attenuation. The resulting conditional response is then rescaled to plant and fleet power.
The resulting leakage bounds should therefore be interpreted as screening-level source-term constraints rather than as facility-specific shielding calculations.

As a historical analogue for neutron-driven atmospheric radiocarbon production, Sakharov and Pauling treated radiocarbon from above-ground nuclear testing as a global cumulative source term. In 1958, Sakharov and, shortly thereafter, Pauling argued that radiocarbon produced by atmospheric nuclear explosions should be evaluated as a long-lived global burden: \textsuperscript{14}C enters the carbon cycle and is incorporated efficiently into living matter~\cite{Sakharov1958,Pauling1958}. Fusion energy production differs fundamentally from weapons testing in purpose, source geometry and operational context, but the radiochemical pathway is shared. Escaped neutrons entering atmospheric nitrogen can produce globally mixed \textsuperscript{14}C; for deployment-scale technologies, cumulative source strength, biospheric uptake, atmospheric mixing, and fleet size therefore become central quantities alongside prompt dose and local activation. Natural \textsuperscript{14}C production provides a useful benchmark. Cosmogenic interactions produce radiocarbon in the atmosphere at an average global rate of about $2~\mathrm{atoms~cm^{-2}~s^{-1}}$~\cite{Lal1967}, corresponding to a total source term of roughly $7.5~\mathrm{kg~yr^{-1}}$, or $1\times10^{19}~\mathrm{atoms~s^{-1}}$~\cite{Usoskin2012}. We use 10\% of the natural source term as an illustrative reference level that is comparable to ordinary natural variability, policy-relevant, and easy to rescale.

\section*{Results and analysis}

\subsection*{Representative atmospheric $^{14}$C yields from leaked neutrons}

To estimate the atmospheric radiocarbon yield per leaked neutron, we performed MCNP\textsuperscript{\textregistered}6.2 transport calculations for isotropic sources in dry air above a planar ground boundary (Methods and Supplementary Table~1)~\cite{Goorley2012}. 
The source in these calculations should be interpreted as a terminal leakage source, i.e., a neutron current 
that has already passed through the reactor-specific shielding, port, duct, window, or containment system and 
entered nitrogen-containing gas. The calculations therefore estimate the atmospheric $^{14}$C response conditional on 
such leakage. 
Cases were run for 14.1~MeV, 10~keV, and thermal source neutrons, and for several representative ground materials. The purpose was not to construct a site-specific licensing model, but to establish the order of magnitude of the net probability that a leaked neutron entering air ultimately produces $^{14}$C through the $^{14}$N$(n,p)^{14}$C nuclear reaction.

For 14.1~MeV source neutrons, the net conversion probability is 0.506 in an ``air only'' case and 0.25--0.38 for near-ground cases; for fleet-scale estimates we use the representative value $P_{^{14}\mathrm{C}}\gtrsim 0.33$. The lower yield for fast leakage reflects competing inelastic reactions on nitrogen and oxygen before full moderation. Once neutrons enter air in the epithermal or thermal range, capture on nitrogen dominates and the net yield rises to roughly 0.5--0.95~\cite{ENDFVIII,Mughabghab2018,TorresSanchez2023}. (The upper bound matches the absorption 
branching ratio for $^{14}$N$(n,p)^{14}$C relative to competing absorption processes in air.) The key point is that leakage into nitrogen-containing gas is far from benign. Within the representative geometries considered here, uncertainties associated with local surroundings (humidity, altitude, and local composition of air and ground material) change the inferred leakage requirement by factors of order unity, not by orders of magnitude, and are not expected to alter the central scaling qualitatively.

For a 1~GWe deuterium--tritium plant, corresponding to about 3~GW$_\mathrm{th}$ at 33\% conversion efficiency, the neutron source strength is approximately \SI{1.1e21}{s^{-1}}. 
If we define $f_{\mathrm{leak}}$ as 
the effective fraction of source neutrons that enter nitrogen-containing gas after all shielding, absorbers, port geometries, ducts, windows, buildings, and engineered mitigation measures, then the atmospheric radiocarbon production rate is approximately
\[
R_{^{14}\mathrm{C}} \simeq 1.1\times10^{21} f_{\mathrm{leak}} P_{^{14}\mathrm{C}}\ \text{atoms s}^{-1}.
\]
Using $P_{^{14}\mathrm{C}}=0.33$ gives $R_{^{14}\mathrm{C}}\simeq 3.6\times10^{20}f_{\mathrm{leak}}$ atoms s$^{-1}$. On that basis, 1\% total neutron leakage from a 1~GWe plant would yield about $3.6\times10^{18}$ $^{14}$C atoms s$^{-1}$, roughly one-third of the natural global production rate. The result applies to fusion concepts whose fuel cycles produce substantial neutron fluxes and whose leaked neutrons can enter nitrogen-containing gas. It does not apply in the same way to genuinely aneutronic concepts, although those have much more demanding confinement and plasma-performance requirements~\cite{Nevins1998}.

\subsection*{From plant-scale leakage to a fleet-scale design constraint}

\begin{figure*}[t]
\centering
\includegraphics[width=\textwidth]{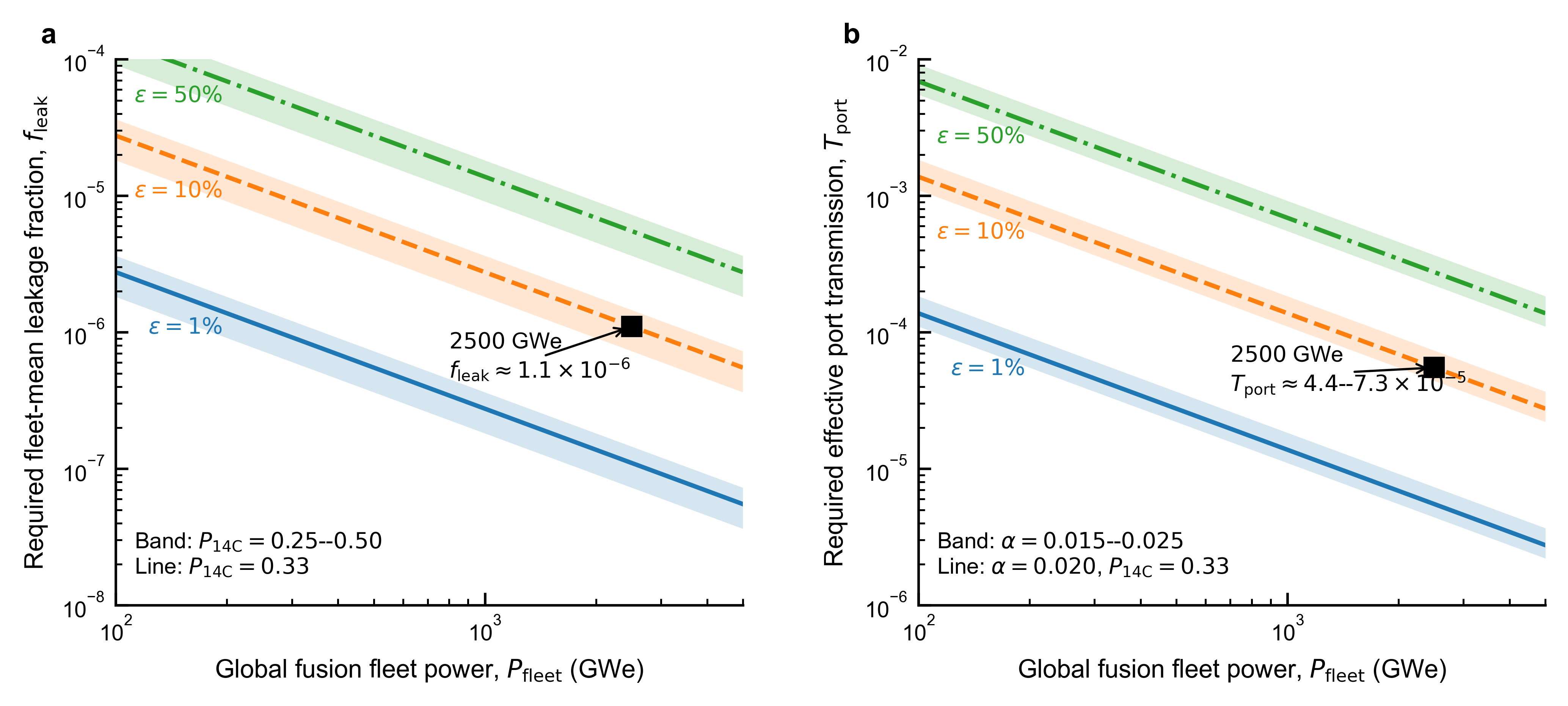}
\caption{
Scaling of representative neutron-leakage requirements for neutron-producing fusion energy systems.
\textbf{a,} Required fleet-mean leakage fraction, $f_{\rm leak}$, as a function of global fusion fleet power, $P_{\rm fleet}$, for allowed anthropogenic $^{14}$C source terms $\epsilon \equiv R^{\rm fusion}_{14{\rm C}}/R^{\rm nat}_{14{\rm C}}$ of 1\%, 10\%, and 50\%. Solid, dashed, and dash-dotted lines denote the three values of $\epsilon$; shaded bands show the effect of representative fast-neutron atmospheric conversion probabilities $P_{14{\rm C}}=0.25$--0.50, with the central lines evaluated at the central illustrative value $P_{14{\rm C}}=0.33$. The square marks the illustrative case $P_{\rm fleet}=2500$~GWe and $\epsilon=10\%$, for which $f_{\rm leak}\approx 1.1\times10^{-6}$.
\textbf{b,} Corresponding effective port-transmission requirement, $T_{\rm port}$, for a port-dominated inertial-fusion-energy leakage model, $f_{\rm leak}\simeq \alpha T_{\rm port}$, where $\alpha$ is the total port solid-angle fraction. Shaded bands show $\alpha=0.015$--0.025, with central lines evaluated at $\alpha=0.020$ and $P_{14{\rm C}}=0.33$. The square marks the same illustrative case, for which $T_{\rm port}\sim (4.4$--$7.3)\times10^{-5}$.
The figure is intended as a representative scaling summary, not as a site-specific shielding calculation.
}
\label{fig:scaling}
\end{figure*}

The constraint becomes sharper when fusion is considered as infrastructure rather than as an isolated facility. For a worldwide fusion fleet with electric output $P_{\mathrm{fleet}}$, the mean leakage fraction required to keep fusion-derived atmospheric radiocarbon production below a fraction $\epsilon$ of the natural source term is approximately
\[
f_{\mathrm{leak}} \lesssim 1.1\times10^{-6}
\left(\frac{\epsilon}{0.1}\right)
\left(\frac{2500\,\mathrm{GWe}}{P_{\mathrm{fleet}}}\right)
\left(\frac{0.33}{P_{^{14}\mathrm{C}}}\right).
\]
For a 2500~GWe fleet and the 10\% benchmark, the required fleet-average leakage fraction is therefore of order $10^{-6}$. The same scaling gives values of order $10^{-7}$, $10^{-6}$, and $5\times10^{-6}$ for 1\%, 10\%, and 50\% benchmarks, respectively, at the same fleet size (Supplementary Table~2). Those values are indicative design targets, and they scale linearly with the chosen benchmark, fleet size, and per-neutron conversion probability. Figure~\ref{fig:scaling} summarizes this scaling and shows how the corresponding fleet-level leakage requirement maps onto an effective transmission requirement for port-dominated inertial-fusion concepts.

These trends underscore two points. First, the ppm-level leakage requirement is not tied to a single benchmark, but emerges across a range of deployment scales and tolerable perturbations to the natural $^{14}$C source term. Second, if leakage is dominated by optically required penetrations, the fleet-level requirement maps onto stringent effective transmission limits for individual IFE ports.

This constraint is not a substitute for prompt-dose shielding requirements, and in many facility regions prompt neutron and gamma dose limits will be the more restrictive design driver. The atmospheric $^{14}$C constraint matters because it is distinct from maximally exposed individual or site-boundary dose constraints. Prompt-dose limits control dose rates at occupied locations and facility boundaries, whereas atmospheric radiocarbon depends on the integrated neutron current that enters nitrogen-containing gas and is ultimately released or mixed into the environment. The distinction is relevant for streaming paths through ports, ducts, beamlines, roof or wall penetrations, shielded but ventilated volumes, and other regions where occupancy may be low or local dose may be controlled but neutron capture in air can still contribute to a persistent global radiocarbon inventory. Conversely, if a design demonstrates that leaked neutrons were absorbed before entering nitrogen-containing gas, or if that activated gas were retained and not released, the atmospheric $^{14}$C constraint would be correspondingly reduced. The issue is therefore not an alternative to conventional shielding analysis, but an additional source-term accounting requirement for neutron-producing fusion systems at scale.

For scale, a sustained anthropogenic source equal to the natural global
$^{14}$C production rate would correspond, in equilibrium, to an
incremental internal dose of order \SI{10}{\micro\sievert\,y^{-1}}
(Supplementary Note~3)~\cite{Bharath2021,ICRP103,Masuda2020,ICRP158}.
This dose is small by ordinary individual-dose standards, but the source
term is diffuse, weakly attributable, and long-lived. Atmospheric
$^{14}$C is therefore best understood as a cumulative global inventory
perturbation, i.e., an energy-system externality that matters primarily at
scale, even when the contribution from any one facility is modest.

\subsection*{Architecture, mitigation, and governance}

Fusion pathways face different radiocarbon-control burdens. Magnetic fusion energy can in principle exploit largely continuous metallic and concrete shielding, although heating, pumping, and diagnostic penetrations remain important streaming paths~\cite{Santoro1998}. Laser inertial fusion energy (IFE) is a sharper case because the reactor must preserve direct optical access to the target. (Magnetic inertial fusion concepts require vacuum conducting paths and have similar challenges.) Existing and proposed laser-driven systems use tens to hundreds of beams delivered through multiple ports~\cite{Wilhelmsen2011,Bayramian2011,Meier2013_FED}. For aggregate port solid-angle fractions $\alpha\sim0.015$--0.025, a plant-wide leakage target below $10^{-6}$ implies effective port transmission below roughly $(4$--$7)\times10^{-5}$. 
In the port-dominated limit, $f_{\mathrm{leak}} \simeq (1-\alpha)T_{\mathrm{bulk}} + \alpha T_{\mathrm{port}} \simeq \alpha T_{\mathrm{port}}$, so this corresponds to port attenuation factors of order $1.5 \times 10^4$--$2.5 \times 10^{4}$, or 
$>99.99\%$ neutron attenuation, maintained over multi-decade operation at multi-Hz repetition rates.
That level of attenuation is not obviously unattainable, but it would have to be delivered while preserving optical throughput, repetition rate, maintainability, and lifetime under severe radiation and debris loading. The practical implication is that atmospheric radiocarbon should enter IFE trade studies early, not after major architectural choices are fixed.

Several mitigation options merit early engineering study. They include nitrogen-depleted atmospheres in relevant plant volumes, neutron-absorbing optical materials, beam labyrinths that interrupt line-of-sight streaming, subsurface siting, and downstream carbon-capture approaches~\cite{TaylorCortes2014,Guillemet2025,Dosovitskiy2020,Singh2016GdGlass,Colangeli2021,INSAG10}.
Each option carries coupled engineering constraints: nitrogen-depleted atmospheres imply large conditioned volumes and oxygen-deficiency controls; neutron-absorbing optical elements must maintain high optical quality under neutron,
gamma, and debris loading; labyrinths and shutters can compromise repetition rate, alignment, and optical path length; subsurface
siting does not eliminate production in nitrogen-filled beamlines or ports; and downstream processing must address gaseous $^{14}$CO$_2$, not only particulates. 
For some architectures these mitigation options are likely to be first-order design variables rather than peripheral add-ons, especially if very low leakage is required over multi-decades of operation.

The policy implication is similarly important. Existing nuclear regulation already considers environmental pathways, and licensing analyses do not stop at the site boundary. Even so, frameworks such as 10~CFR Part~20 remain organized primarily around local public dose and the maximally exposed individual~\cite{10CFR20}. A globally distributed, weakly attributable $^{14}$C source fits awkwardly within that facility-centered logic. Earlier fusion safety studies, including HYLIFE-II, largely treated neutron leakage as a prompt-dose and activation issue rather than as a contributor to the global radiocarbon inventory~\cite{MoirHYLIFEII1994}. 

If neutron-producing fusion moves toward wide deployment, atmospheric radiocarbon should be addressed explicitly in 
environmental assessment, design targets, technology comparisons, fleet-level source accounting, and long-term
monitoring. Public-boundary standoff can reduce prompt dose, but it does not materially relax the $^{14}$C constraint once
neutrons have escaped into nitrogen-containing gas. 

\section*{Discussion}

Atmospheric radiocarbon should be treated as a deployment-scale source term for neutron-producing fusion. 
Our transport calculations show that a leaked 14.1~MeV neutron entering nitrogen-containing air can generate $^{14}$C with probability of order 0.25--0.50 under representative near-ground conditions, with larger yields possible for softer leakage spectra. This conversion probability makes small leakage fractions consequential at energy-system scale. A single 1~GWe plant with percent-level atmospheric neutron leakage can approach the natural global $^{14}$C source to within factors of a few. For the representative conversion probabilities used here, a 2500~GWe fusion fleet constrained to 10\% of natural production requires a mean atmospheric leakage fraction of order $10^{-6}$. The result is therefore an infrastructure-scale design target.

The Sakharov--Pauling history sharpens this framing. Above-ground nuclear testing showed that neutron production in air can create a globally mixed, long-lived radiocarbon burden~\cite{Sakharov1958,Pauling1958}. The relevance to fusion is that escaped neutrons encountering atmospheric nitrogen drive the same nuclear reaction, and thus the resulting $^{14}$C enters the same atmospheric and biospheric reservoirs. Fusion energy differs from weapons testing in purpose, source geometry, operating conditions, and governance, though they share the same source-term pathway. That pathway shifts the assessment beyond plant-bound 
quantities---prompt dose, local activation, and component damage---to cumulative source strength, atmospheric mixing, carbon-cycle uptake, and fleet-scale source-term accounting.

These calculations provide a screening analysis rather than a site-specific licensing model. Real facilities will have complex shielding, penetrations, buildings, ducts, beamlines, exhaust systems, surrounding terrain, and time-varying meteorology. Those details may change the atmospheric yield for a particular design and site but they do not remove the source-term constraint because the fleet-scale result scales linearly with fusion power, leakage fraction, and per-neutron conversion probability. The appropriate response is therefore to identify where neutrons can enter nitrogen-containing gas, quantify the leakage spectrum and solid angle, and design the shielding, absorbers, port geometries, and non-nitrogen buffer environments to suppress the atmospheric source.

The practical implication is that atmospheric $^{14}$C should be included early in fusion environmental assessment and design optimization, especially for architectures with optically open ports, beamlines, or other direct neutron-leakage pathways. Future studies should report atmospheric neutron-leakage fractions, leakage spectra, assumed $^{14}$C yields, and fleet-scale source-term comparisons against natural production. Fusion can remain a low-carbon energy option while carrying this constraint explicitly. Its environmental case is stronger when global, long-lived source terms are quantified before deployment decisions lock in the architecture.

\section*{Methods}

\subsection*{Monte Carlo transport model}

The transport model was designed to estimate a conditional atmospheric response: the number of $^{14}$N$(n,p)^{14}$C reactions in air per neutron that has already leaked into nitrogen-containing gas. It was not intended to predict the leakage fraction from any particular reactor, port, duct, shield, building, or site.

Atmospheric $^{14}$C production probabilities were estimated with the Monte Carlo N-Particle transport code MCNP\textsuperscript{\textregistered} version 6.2~\cite{Goorley2012}. The representative dry-ground input deck used analog neutron transport (\texttt{mode n}) with $10^6$ source histories per case and neutron interaction data from \texttt{.80c} tables~\cite{ENDFVIII}. The model geometry comprised a small spherical source region embedded in dry air above a planar ground surface at $z=0$, with an air half-space above and a ground half-space below. In the representative 10-m dry-ground case, the source sphere had radius 5~cm and was centered at $(0,0,1000\ \mathrm{cm})$. A large cylindrical outer boundary of radius $2.0\times10^7$~cm and planes at $z=\pm 2.0\times10^7$~cm enclosed the problem, with particles terminated in an exterior graveyard cell.

Dry air in the representative deck was modeled at density $1.205\times10^{-3}$~g~cm$^{-3}$ with atomic fractions N 0.755268, O 0.231781, and Ar 0.012951. The representative dry-soil half-space was modeled at density 1.80~g~cm$^{-3}$ with atomic fractions H 0.020, O 0.580, Na 0.015, C 0.020, Al 0.070, Si 0.250, K 0.015, Ca 0.020, and Fe 0.010. The baseline source was isotropic and monoenergetic at 14.1~MeV. Additional cases for source heights of 1, 10, and 100~m, together with an ``air only'' limit and softer source energies of 10~keV and 0.0253~eV, were generated by analogous modifications to source height, source energy, and ground material cards. Wet-ground cases were implemented as dry ground with 2.8 wt\% added water, using the same geometry and source definitions as the corresponding dry-ground case.

\subsection*{Tallies and derived quantities}

The principal quantity of interest was the expected number of $^{14}$N$(n,p)^{14}$C reactions in air per source neutron. In the representative deck this quantity was scored in the air cell with an F4 cell-flux tally coupled to an FM multiplier using MT~=~103 on $^{14}$N, with \texttt{SD24 1} normalization. Auxiliary tallies recorded neutron flux in air and ground, total neutron removal in air and ground, boundary currents through the radial, top, and bottom surfaces, and an energy-binned leakage spectrum spanning $10^{-8}$ to 20~MeV. The representative deck also used a neutron cutoff of $10^{-9}$~MeV and the \texttt{phys:n j 20} setting from the supplied input. The net conversion probabilities reported in Supplementary Table~1 are therefore the tallied $^{14}$N$(n,p)^{14}$C reaction rates in air per source neutron for each case.

The calculations were generated from a common input template, with case-specific changes to source energy, source height, and ground material. The full case matrix and the resulting $^{14}$C-production probabilities are reported in Supplementary Table~1. Supplementary Table~1 also reports characteristic confidence intervals for the corresponding MCNP tallies. 
These statistical uncertainties quantify Monte Carlo sampling error only and do not include systematic uncertainty associated with atmospheric composition, humidity, terrain, facility geometry, structural materials, or time-varying meteorology. We therefore use the calculated probabilities as representative atmospheric yields for scaling analysis rather than as site-specific licensing estimates.

\subsection*{Plant- and fleet-scale source-term scaling}

The plant- and fleet-scale estimates use the tallied reaction probability $P_{^{14}\mathrm{C}}$ as a response function for terminal leakage. For a deuterium--tritium plant producing 1~GWe, corresponding to approximately 3~GW$_\mathrm{th}$ at 33\% thermal-to-electric conversion efficiency, the fusion-neutron source strength is approximately $\dot N_n\simeq1.1\times10^{21}~\mathrm{s^{-1}}$. If $f_{\mathrm{leak}}$ is the effective fraction of source neutrons that enter nitrogen-containing gas after all shielding and mitigation measures, the atmospheric carbon-14 production rate is $R_{^{14}\mathrm{C}}\simeq \dot N_n f_{\mathrm{leak}} P_{^{14}\mathrm{C}}$. The fleet-scale leakage bounds follow by requiring this source term, multiplied by total fleet electric power in GWe, to remain below a chosen fraction $\epsilon$ of the natural global carbon-14 production rate. These scalings use no facility-specific leakage estimate.

\section*{Data availability}

The numerical values used to generate Fig.~\ref{fig:scaling} are given by the equations in the text and the values tabulated in Supplementary Tables~1 and~2. The MCNP\textsuperscript{\textregistered} input decks are subject to LANL/DOE/NNSA release review.

\section*{Code availability}

The simulations in this study were performed using the Monte Carlo N-Particle\textsuperscript{\textregistered} code version 6.2 (MCNP\textsuperscript{\textregistered}6.2). MCNP\textsuperscript{\textregistered}6.2 is a proprietary software package developed by Los Alamos National Laboratory and distributed by the Radiation Safety Information Computational Center (RSICC) at Oak Ridge National Laboratory. Due to export control regulations and licensing restrictions, the MCNP\textsuperscript{\textregistered}6.2 source code and executables cannot be made publicly available by the authors; access requests must be directed to the RSICC software catalog. No custom supplemental scripts or post-processing codes were used. The specific MCNP\textsuperscript{\textregistered}6.2 input files (including geometry, material specifications, and tally configurations) generated and used to produce the results in this paper are available upon reasonable request from the corresponding author 
and will be deposited in a persistent public repository if the manuscript proceeds to publication, 
subject to institutional release review. 

\section*{Acknowledgements}

This work was performed under the auspices of the U.S. Department of Energy by Triad National Security, LLC, Los Alamos National Laboratory, for the U.S. Department of Energy's National Nuclear Security Administration under
Contract No. 89233218CNA000001.
Los Alamos Unlimited Release number: LA-UR-26-24308. Approved for public release; distribution is unlimited.
The authors thank Drs. John Kline, Todd Urbatsch, William Daughton, Andrei Simakov, Joseph Smidt, John Scott, David Meyerhofer, and Mark Chadwick for useful discussions.

\section*{Author Contributions}
B.A. and J.M.-S. conceived the study and developed the central results. B.A. designed and executed the code calculations and 
wrote the initial draft. B.A. and J.M-S. discussed the results, contributed to the interpretation of the data, and 
edited the final manuscript.

\section*{Competing interests}

The authors declare no competing interests.

\section*{Generative AI statement}

Portions of this manuscript were prepared with the assistance of the large language models Claude (Anthropic) and ChatGPT (OpenAI) for language editing and organizational refinement. No AI was used for calculations or data generation. 
The authors take full responsibility for the content.

\clearpage
\section*{Supplementary Information}

For review purposes, the supplementary material is included below in the same document. The notes repeat the source-energy matrix, tally interpretation, and scaling equations needed to reproduce the main order-of-magnitude estimates without relying on any reactor-specific leakage calculation.

\subsection*{Supplementary Note 1 | Representative atmospheric radiocarbon-production cases}

The central quantity used in the main text is the net probability that a leaked neutron entering nitrogen-containing gas ultimately produces radiocarbon through
$^{14}\mathrm{N}(n,p)^{14}\mathrm{C}$. We estimated this probability over source energies relevant to fusion-neutron leakage using neutron-transport calculations with MCNP\textsuperscript{\textregistered} version 6.2~\cite{Goorley2012} and standard evaluated neutron-interaction data~\cite{ENDFVIII,Mughabghab2018,TorresSanchez2023}. The calculations considered monoenergetic neutron sources at heights of 100~m, 10~m, and 1~m above ground, together with an ``air only'' case approximating a source far from any surface. Ground materials were selected to span representative moderation and absorption behavior: dry ground, wet ground, concrete, and water. The three source energies were 14.1~MeV, 10~keV, and 0.0253~eV, corresponding respectively to primary deuterium--tritium fusion neutrons, a representative epithermal leakage energy, and thermal neutrons.

{\small\textbf{Supplementary Table 1 |} Net carbon-14 breeding fractions $P_{^{14}\mathrm{C}}^{14.1}$, $P_{^{14}\mathrm{C}}^{0.01}$, and $P_{^{14}\mathrm{C}}^{\mathrm{th}}$ for source-neutron energies of 14.1~MeV, 10~keV, and 0.0253~eV, respectively, by ground type and source height. Wet ground is dry ground with 2.8 wt\% added water. The tabulated values are rounded to three decimal places. The MCNP statistical relative errors for the underlying tallies were typically $R\simeq(1$--$2)\times10^{-3}$, corresponding to characteristic 95\% statistical confidence intervals with half-widths of order $0.3\%$ of the reported probability, or approximately $(1$--$3)\times10^{-3}$ in absolute probability over the range shown. These intervals reflect Monte Carlo counting statistics only and do not include model, geometry, material-composition, or nuclear-data uncertainty.
\\[0.4em]
\begin{center}
\begin{tabular}{lcccc}
\toprule
Ground type & Source height & $P_{^{14}\mathrm{C}}^{14.1}$ & $P_{^{14}\mathrm{C}}^{0.01}$ & $P_{^{14}\mathrm{C}}^{\mathrm{th}}$ \\
\midrule
air only   & $\infty$ & 0.506 & 0.953 & 0.947 \\
dry ground & 100 m & 0.376 & 0.729 & 0.930 \\
dry ground & 10 m  & 0.358 & 0.560 & 0.723 \\
dry ground & 1 m   & 0.330 & 0.542 & 0.639 \\
wet ground & 10 m  & 0.318 & 0.531 & 0.736 \\
concrete   & 10 m  & 0.339 & 0.565 & 0.710 \\
water      & 10 m  & 0.250 & 0.502 & 0.762 \\
\bottomrule
\end{tabular}
\end{center}
}

The calculations show that the net radiocarbon yield depends strongly on the leakage spectrum. For 14.1-MeV source neutrons, competing fast-neutron reactions on nitrogen and oxygen reduce the probability of eventual $^{14}$C production to $P_{^{14}\mathrm{C}}\approx0.25$--0.51 in the representative cases considered here. Once neutrons enter air in the epithermal or thermal range, however, the dominant absorption channel in dry air is $^{14}\mathrm{N}(n,p)^{14}\mathrm{C}$, and the net yield rises to approximately 0.5--0.95, depending on the surrounding materials and source height. Moderation without capture is therefore not, by itself, a mitigation mechanism for atmospheric radiocarbon production. From the standpoint of atmospheric $^{14}$C control, the relevant objective is attenuation or absorption before neutrons leak into nitrogen-containing gas.

The numerical values in Supplementary Table~1 are geometry-dependent and should be interpreted as representative rather than universal. The calculations do not include building structures, terrain, or time-varying atmospheric composition. Increased humidity would tend to enhance moderation, whereas altitude and lower-density air would increase neutron transport lengths. These effects may shift the detailed probabilities, but they are not expected to change the main conclusion: the per-neutron probability of atmospheric radiocarbon production can be large enough that neutron-leakage control is relevant at deployment scale.

\subsection*{Supplementary Note 2 | Scaling to fleet-average leakage bounds}

For a deuterium--tritium fusion plant producing 1~GWe, corresponding to approximately 3~GW$_\mathrm{th}$ at 33\% conversion efficiency, the fusion-neutron source strength is about
\[
\dot N_n \simeq 1.1\times10^{21}~\mathrm{s^{-1}}.
\]
If a fraction $f_{\mathrm{leak}}$ of those neutrons leaks into air, the resulting atmospheric radiocarbon production rate is
\[
R_{^{14}\mathrm{C}} \simeq \dot N_n f_{\mathrm{leak}} P_{^{14}\mathrm{C}} \simeq 1.1\times10^{21} f_{\mathrm{leak}} P_{^{14}\mathrm{C}}~\mathrm{atoms~s^{-1}}.
\]
Using the low-end representative value $P_{^{14}\mathrm{C}}=0.33$ gives
\[
R_{^{14}\mathrm{C}} \simeq 3.6\times10^{20} f_{\mathrm{leak}}~\mathrm{atoms~s^{-1}}.
\]

For a worldwide fusion fleet with total electrical output $P_{\mathrm{fleet}}$, and allowing an anthropogenic contribution equal to a fraction $\epsilon$ of the natural radiocarbon source term, the mean leakage requirement is
\[
 f_{\mathrm{leak}} \lesssim 1.1\times10^{-6}
 \left(\frac{\epsilon}{0.1}\right)
 \left(\frac{2500~\mathrm{GWe}}{P_{\mathrm{fleet}}}\right)
 \left(\frac{0.33}{P_{^{14}\mathrm{C}}}\right).
\]
This expression makes explicit that the central result is not tied to a single benchmark choice.

\begin{center}
\small\textbf{Supplementary Table 2 |} Illustrative fleet-average leakage bounds for a 2500~GWe fleet, assuming $P_{^{14}\mathrm{C}}=0.33$.\\[0.4em]
\begin{tabular}{ccc}
\toprule
Allowed fusion-derived share of natural $^{14}$C source term, $\epsilon$ & Benchmark interpretation & Required mean $f_{\mathrm{leak}}$ \\
\midrule
0.01 & 1\% of natural production & $1.1\times10^{-7}$ \\
0.10 & 10\% of natural production & $1.1\times10^{-6}$ \\
0.50 & 50\% of natural production & $5.5\times10^{-6}$ \\
\bottomrule
\end{tabular}
\end{center}

Fleet heterogeneity, staged deployment, and learning-by-doing would all matter in practice. The simple scaling above should therefore be read as a deployment thought experiment that identifies the order of magnitude of the leakage target implied by sustained, large-scale deployment. It is not a forecast of any one reactor design, regulatory regime, or commercialization pathway.

\subsection*{Supplementary Note 3 | Dose scaling from increased radiocarbon levels}

A sustained doubling of ambient $^{14}$C specific activity would increase the internal dose from naturally incorporated radiocarbon by approximately a factor of two, from a present-day value of order $12~\mu\mathrm{Sv\,y^{-1}}$ to roughly $24~\mu\mathrm{Sv\,y^{-1}}$, implying an incremental dose of
\[
\Delta E \sim 12~\mu\mathrm{Sv\,y^{-1}}.
\]
Under the simplifying assumption that dose scales linearly with environmental $^{14}$C production, and adopting the nominal low-dose, low-linear-energy-transfer stochastic detriment coefficient
\[
r \approx 5\times10^{-2}~\mathrm{Sv^{-1}},
\]
the associated incremental annual risk is of order
\[
\Delta R_{\mathrm{yr}} \sim r\,\Delta E \sim 6\times10^{-7},
\]
while the corresponding lifetime increment over 70 years is
\[
\Delta R_{\mathrm{life}} \sim r\,(70\,\Delta E) \sim 4\times10^{-5}.
\]
These estimates should be regarded as order-of-magnitude values, since they depend on the extent to which elevated $^{14}$C production is sustained long enough to raise biospheric specific activity, as well as on uncertainties in carbon-cycle turnover, tissue retention, and dietary dosimetry~\cite{Bharath2021,ICRP103,Masuda2020,ICRP158}. Their relevance here is not that the corresponding individual doses are large in the ordinary regulatory sense. Rather, they illustrate why a persistent atmospheric $^{14}$C source is best understood as a diffuse and long-lived collective burden rather than as a conventional prompt-dose issue.

\end{document}